\documentclass[letterpaper, 10 pt, conference]{ieeeconf}  % Comment this line out if you need a4paper

\usepackage{graphicx}  %added
\usepackage{epstopdf}  %added
\usepackage{subfig}    %repalce sunfigure with subfloat
\usepackage{amsmath}   %added
\usepackage{boxhandler}%added by ali
\usepackage{dsfont}    %added by ali
\usepackage[flushleft]{threeparttable} %added by ali
\let\labelindent\relax %added to use enumitem; disables the IEEE format for laleintent
\usepackage{enumitem}  %added by ali
\usepackage{booktabs}  %added by ali
\usepackage{multirow}  %added by ali
\usepackage{url} 	   %added by ali
\usepackage{setspace}  %added by ali
\usepackage{textcomp}  %added by ali
\usepackage{color}     %added by ali
\usepackage{booktabs} %added by ali
\usepackage{array}    %added by ali
\usepackage{caption}  %added by ali
\usepackage[font=footnotesize]{caption}
\usepackage{epsfig}   %1st paper
\usepackage{mathptm}  %1st paper
\usepackage{amssymb} 	%1st paper
\usepackage{amsxtra}    %1st paper   
\usepackage{verbatim}  %1st paper
\usepackage{graphics} %1st paper
\usepackage{pdflscape} %1st paper
\usepackage{enumerate}
\usepackage{enumitem}
\usepackage{setspace}
\usepackage[export]{adjustbox}
\usepackage{setspace}
\usepackage{bm}
\usepackage{lipsum}% http://ctan.org/pkg/lipsum
\usepackage{algorithm}
\usepackage[noend]{algpseudocode}
\usepackage[noadjust]{cite} %to group citations

\IEEEoverridecommandlockouts                              % This command is only needed if 
\newcommand*{\TitleFont}{%
       \usefont{\encodingdefault}{\rmdefault}{m}{n}%
       \fontsize{24}{29}%
       \selectfont}
       
\begin{document}      
     
% key words: Mixed-integer linear programming; Autonomous vehicle-intersection coordiantion; Vehicle arrival scheduling; Traffic microsimulation model; Connected vehicles; intelligent traffic systems 

\title{\TitleFont
Multi-Intersection Traffic Management for Autonomous Vehicles via Distributed Mixed Integer Linear Programming
}

\author{Faraz Ashtiani \quad S. Alireza Fayazi \quad Ardalan Vahidi% <-this % stops a space
% <-this % stops a space
%
\thanks{Faraz Ashtiani ({\tt\footnotesize fashtia@g.clemson.edu}), Seyed Ali Reza Fayazi ({\tt\footnotesize sfayazi@g.clemson.edu}), and Ardalan Vahidi ({\tt\footnotesize avahidi@clemson.edu}) are with the Department of Mechanical Engineering, Clemson University, Clemson, SC 29634-0921, USA. }%
}

\maketitle
\thispagestyle{empty}
\pagestyle{empty}

%%%%%%%%%%%%%%%%%%%%%%%%%%%%%%%%%%%%%%%%%%%%%%%%%%%%%%%%%%%%%%%%%%%%%%%%%%%%%%%%
\begin{abstract}
This paper extends our previous work in \cite{fayazi2017ACC, fayazi2017CCTA}, on optimal scheduling of autonomous vehicle arrivals at intersections, from one to a grid of intersections. A scalable distributed Mixed Integer Linear Program (MILP) is devised that solves the scheduling problem for a grid of intersections. A computational control node is allocated to each intersection and regularly receives position and velocity information from subscribed vehicles. Each node assigns an intersection access time to every subscribed vehicle by solving a local MILP. Neighboring intersections will coordinate with each other in real-time by sharing their solutions for vehicles' access times with each other. Our proposed approach is applied to a grid of nine intersections and its positive impact on traffic flow and vehicles' fuel economy is demonstrated in comparison to conventional intersection control scenarios.
\end{abstract}

%%%%%%%%%%%%%%%%%%%%%%%%%%%%%%%%%%%%%%%%%%%%%%%%%%%%%%%%%%%%%%%%%%%%%%%%%%%%%%%%

%===========================================================================
%  ===========================================================================
\section{Introduction}
Enhancing traffic flow at signalized intersections has been an important topic in transportation research. New intersection control algorithms and technologies  are always needed to respond to ever growing traffic demand in our cities. Proliferation of Vehicle-to-Everything (V2X) connectivity in recent years could be a significant game changer in intersection control. By acknowledging  capabilities of wireless communication, several research studies have been focused on adaptive traffic signal control based on Vehicle-to-Vehicle (V2V) communication \cite{gradinescu2007adaptive, 55goodall2013traffic} and Vehicle-to-Infrastructure (V2I) communication \cite{rakha2011eco}.
Recent studies have shown that establishing one-way communication from traffic signals to vehicles or from vehicles to signals can harmonize the motion of vehicles so they arrive at the intersection when the signal is green, reducing the number of stops, idling intervals, and energy use \cite{12AsadiB.2011, mandava2009arterial, 101kamal2015traffic}. With reciprocal communication between vehicle and traffic signal, the benefits are expected to be even greater. 

Connected and Autonomous Vehicles (CAV) can play another game changing role and revolutionize intersection management. CAVs can subscribe to an upcoming intersection controller and exchange information with it rather effortlessly and frequently. Autonomous cars can also adjust their arrival times more precisely than a human driver to meet arrival times assigned to them by an intersection controller. In this case and with full market penetration of CAVs, the physical traffic signals can be removed as discussed in several papers \cite{dresner2004multiagent, dresner2008multiagent, kamal2015vehicle, 74tachet2016revisiting, zhang2016optimal, de2010analysis, li2006cooperative}.

% % multi_intersection
The current paper is an extension of the previously presented method by Fayazi and Vahidi in \cite{fayazi2017ACC, fayazi2017CCTA, fayazi2018IEEE} on optimal scheduling of autonomous vehicle arrivals at intersections from one to an urban grid of intersections. The main contribution of \cite{fayazi2017ACC, fayazi2017CCTA, fayazi2018IEEE} is simplifying the complexities in vehicle-intersection coordination by formulating it as a mixed integer linear programming problem. An intersection controller were designed to assign arrival times to vehicles and help them form fast moving platoons. The benefit of using such a MILP-based controller is supervising a large number of subscribing vehicles in real-time. Results in \cite{fayazi2017ACC, fayazi2018IEEE} showed a significant reduction in intersection delay and number of stops while ensuring safe uninterrupted passage of vehicles without compromising the average travel time.

In this paper, we observe that effective passage of a queue of cars requires real-time coordination of neighboring intersection controllers. Therefore, we propose intersections that solve their own optimizations but communicate their decisions with each other in a distributed manner. Placement of controllers on the same backend computational node could make communication of decisions very efficient. In this approach, each intersection processes information from its
neighboring vehicles, optimizes their access times using Mixed Integer Linear Programming (MILP) as in \cite{fayazi2017ACC} and communicates the decisions to subscribed vehicles. Then, each intersection passes on a list of its subscribed vehicles and their allotted access times to its neighboring intersections.

% % organization
The paper is organized as follows: %Section \ref{p2_ch:signalControl_sec:scheduling} introduces the problem in simple words. 
The notations used in this paper as well as the proposed intersection are explained in Section \ref{intersection} and our formulations are presented in  Section \ref{sig_control}. These two sections are summaries of intersection design and formulation presented in \cite{fayazi2017ACC}. Extension to multiple intersections is discussed in Section \ref{multi_int}. In Section \ref{p2_ch:signalControl_sec:milp}, the nature of the MILP solution is shown in a simplified case study with only two intersections. Microsimulation benchmarks and test results are presented for a 3$\times$3 intersection grid in Section \ref{bench} and \ref{p2_ch:signalControl_sec:SIL}, followed by conclusions. 
%  ===========================================================================
\section{Proposed Intersection}
\label{intersection}
\begin{figure}
	\centering
	\includegraphics[width=\columnwidth, trim={0.5cm 3.0cm 8.5cm 0cm},clip]{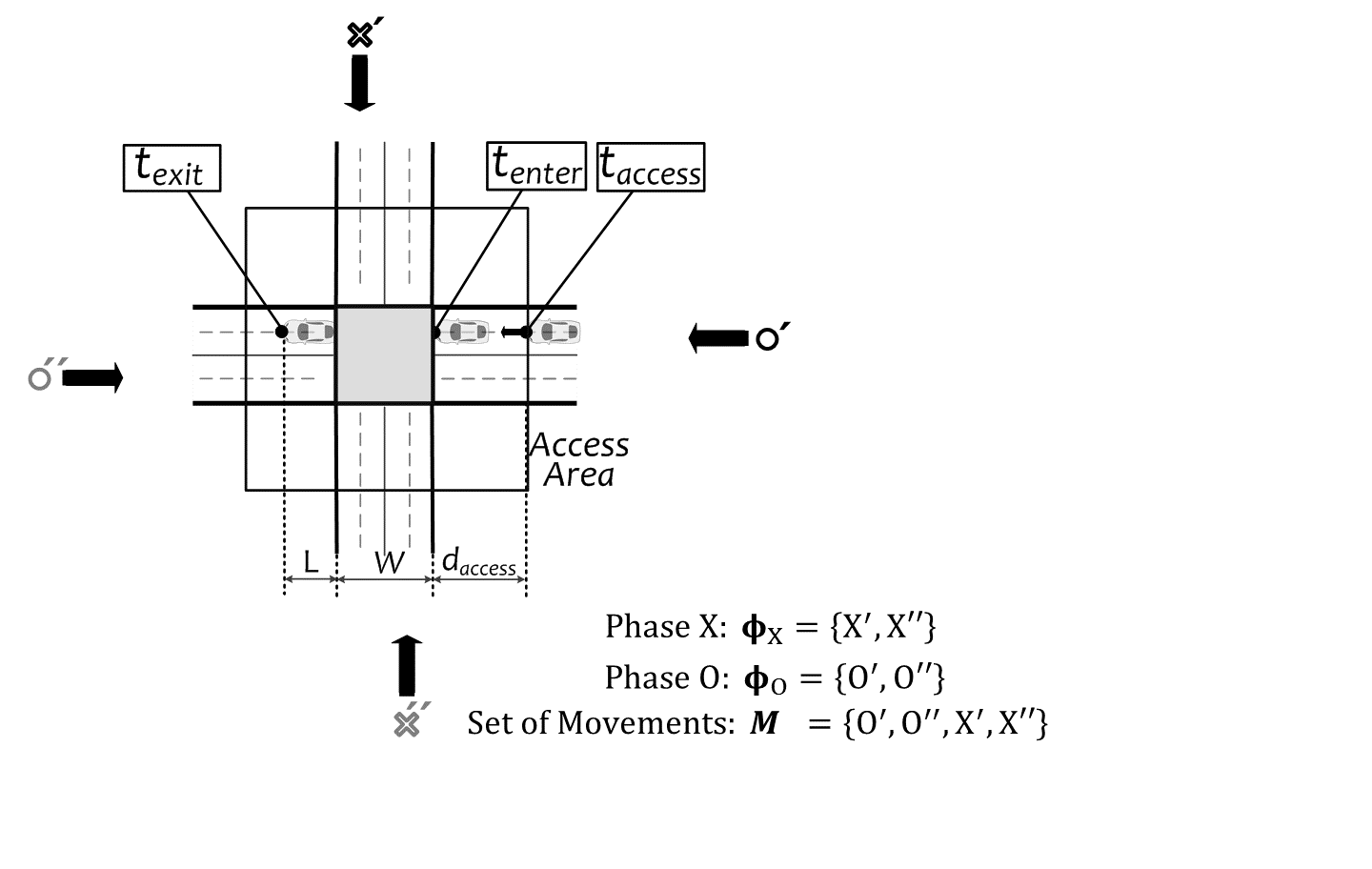}
	\caption{\footnotesize A schematic of phases and regions of the proposed intersection. % (b) in space-time diagram.}
	}
	\label{fig:intersection_system}
	\vspace{-3mm}
	\end{figure}

Ignoring all turns to simplify the presentation of ideas, we assume a square two-phase/four-movement intersection with width $W$=10 m. As shown in Figure \ref{fig:intersection_system}, we consider a two-phase intersection consisting of Phase X and Phase O as $\phi= \{\phi_\textup{X}, \phi_\textup{O}\}$. Each phase %includes 
is allocated to one or more 
%a set of  vehicular traffic movements
non-conflicting movements. The set of all vehicular traffic movements used in this paper is denoted by $M$ (see Figure \ref{fig:intersection_system}). %Each phase includes a set of non-conflicting movements: Phase X ($\phi_\textup{X}=\{\textup{X}',\textup{X}''\}$) corresponds to a set of two south-bound and north-bound traffic movements; similarly, Phase O ($\phi_\textup{O}=\{\textup{O}',\textup{O}''\}$) corresponds to a set of two west-bound and east-bound traffic movements. $M=\{\textup{O}',\textup{O}'',\textup{X}',\textup{X}''\}$ is the set of movements used in this paper (see Figure \ref{fig:intersection_system}). It is assumed that all intersecting roads have the same speed limit denoted by $v_{max}$.
For each intersection, we will assume a subscription process by which the approaching connected vehicles send subscription requests to the intersection control server and announce their presence as well as their intended time of arrival. We represent the list of all subscribed connected vehicles by $CV=\{cv_i\}_{i=1}^n$ where $n$ is the size of $CV$. The list of connected vehicles is sorted by distance to the intersection where $cv_1$ is the closest vehicle to the intersection. %The length of the vehicle is denoted by $L$, and is taken to be 5.0 meter.

For each vehicle approaching an intersection, we are interested in the following time instances: (1) time when the front of the vehicle enters the intersection area at the stop-bar; (2) time when the rear of the vehicle exits the intersection area; (3) time when the front of the vehicle reaches an access distance from the intersection. As shown in Figure \ref{fig:intersection_system}, these time instances are denoted by $t_{enter}$, $t_{exit}$, and $t_{access}$, respectively. In this figure, the intersection and access areas are shown by a shaded area and a solid box, respectively. The access area border is defined by  $d_{access}$ that is the estimated stopping distance of a vehicle in case of a safety concern and is calculated as a function of the road average speed $v_{avg}$. %As it was mentioned above, the safety distance gap denoted by $d_{safety}$ is the estimated distance required to stop the vehicle before crossing the intersection when a dangerous situation is detected. %Using the approach presented in \cite{74tachet2016revisiting}, in the worst-case, this distance gap is equal to the distance required to stop a vehicle traveling at speed limit ($v_{max}$). 
%The safety gap 
%$d_{access}$ is calculated using the kinematic equation (\ref{eqn:dsafety}) as a function of the arterial road average speed $v_{avg}$:
%

The attributes of each vehicle $cv_i \in CV$ ($1\leq i \leq n$) subscribed to an intersection controller are described by:
% %
\begin{equation}
  cv_i = \langle I_i\textup{,}\ m_i\textup{,}\ \phi_i\textup{,}\ d_i\textup{,}\ v_i\textup{,}\ t_{access,i}\textup{,}\ t_{access,des,i}\textup{,}\ t_{exit,i} \rangle
  \label{eqn:cv_i}
\end{equation}
 % %
\noindent where $I_i$ is the intersection ID, $m_i \in M$ is the vehicle movement, $\phi_i \in \phi$ is the phase that $cv_i$'s movement is associated with, $d_i$ is the distance of $cv_i$ to the intersection access point, $v_i$ is the velocity of $cv_i$, $t_{access,i}$ is the assigned time-stamp for $cv_i$ to access the intersection, and $t_{access,des,i}$ is the $cv_i$'s desired access time. Please note that in this paper, we assume that all vehicles prefer to travel at the average velocity $v_{avg}$ %=56.3 kph (35 mph)
and as a result, their distance divided by $v_{avg}$ yields their desired access times with respect to current time ($t_0$=0 sec).

When a vehicle is approaches an intersection, it first sends a subscription request and announces its intended time of arrival. An unsubscribe message is later sent from the vehicle to the intersection controller server at the time the vehicle clears the intersection.
%  ===========================================================================
\section{{MILP}-based Intersection Controller}
\label{sig_control}
\subsection{Objective}
\label{p2_ch:signalControl_sec:formulations_subsec:objective}
The objective of increasing intersection throughput will be formalized here as an optimization problem.
The main goal is to find the optimal sequence and time of arrival ($t_{access}$) for each vehicle such that the difference between the current time ($t_0$) and the expected arrival time of the last vehicle (furthest subscribed vehicle) passing the intersection in a given time window is minimized. This objective is expected to increase the number of vehicles that clear the intersection in a given time:
   		   \begin{equation} 
   		   \begin{split}
   		   & {J_1} = t_{access,n}-t_0 \\ 
   		   & \textbf{\textup{s.t.}} \quad 
   		   n = \#CV \\
   		   & \quad\quad\ t_{access,n} \geq (\{t_{access,1},...,t_{access,n-1}\}) 
   		   \label{eqn:j_1_rephrased_linear}
   		   \end{split}
   		   \end{equation}
   		      	
Minimizing the aforementioned objective %will maximize the number of vehicles that clear the intersection in a given time interval, but it 
could force the vehicles to travel near the speed limit against their preference. To avoid such a scenario, we incorporate the desired arrival time of the vehicles into the optimization problem in such a way that vehicles would not face extreme delay or expedition compared to their desired arrival times. In other words, we  define a cost on the difference between assigned and desired access times for all vehicles:
		\begin{equation}
		{J_2} =  \sum_{i=1}^{n} \lvert t_{access,i} - t_{access,des,i} \rvert\
		\label{eqn:j2}
		\end{equation}
		
The total cost function to be minimized is then:
		\begin{equation}
		J \ = w_1 J_1 \ + \ w_2 J_2
		\label{eqn:jtot}
		\end{equation}

\noindent where $w_1$ and $w_2$ are penalty weights.

\subsection{Constraints}
In this section, a brief description of imposed constraints is presented. For a more in depth explanation regarding paraphrasing the constraints to fit our MILP formulation one can refer to \cite{fayazi2017ACC}.
\label{p2_ch:signalControl_sec:formulations_subsec:constraints}
%Several constraints are imposed to ensure safety. The main challenge is expressing the constraints as a function of access-times so that a linear constrained optimization problem can be derived at the end.
%
\subsubsection{Speed limit and maximum acceleration}
For each vehicle $cv_i$, we should consider the speed and acceleration limit requirement. Since we want to write our constraints dependent on $t_{access}$, we can rephrase speed and acceleration constraints as: 
		\begin{equation}
		 t_{access,i} \geq t_{access,min,i}
		\label{eqn:constraint_accessmin}
		\end{equation}
\noindent where $t_{access,min,i}$ is the earliest time that $cv_i$ can reach the access area.

%74tachet2016revisiting		
		
\subsubsection{Safety gap on the same movement}
Two consecutive vehicles that are traveling on the same movement (e.g. eastbound) should be separated by a safety gap (headway) that is independent of the vehicles' speed to avoid a rear end collision. Headway is defined as time gap between the two vehicles. Therefore, vehicles would be more distant from each other in high velocities.
It is suggested in \cite{74tachet2016revisiting} that a 1 sec headway provides a reasonable upper bound for the response time of an autonomous vehicle. For a standstill vehicle we set $t_{gap1}$ to the time it would take the vehicle body, with length of L, to completely pass over the access point or to 1 sec whichever is larger. For a very slow moving vehicle, we set $t_{gap1}$ to the time it would take a vehicle following directly behind to  decelerate and maintain a minimum safe distance and time headway. To enforce the headway, we add the following constraint on any two consecutive vehicles traveling on the same movement:
	   \begin{equation} 
	   \begin{split}
	   & t_{access,j} - t_{access,k} \geq t_{gap1} \\ 
	   & \ \quad \quad cv_j, cv_k \in CV, \quad d_j \ge d_k ; \\
	   & \ \quad \quad m_j, m_k \in M, \quad m_j=m_k.
	   \label{eqn:constraint_tgap1}
	   \end{split}
	   \end{equation}

\subsubsection{Safety gap on different movements}
Two vehicles traveling on different phases (conflicting movements) also need to be separated by a safety gap. This time gap, if selected properly, guarantees that a vehicle can only enter the access area after all conflicting vehicles have left the intersection area. Considering two vehicles $cv_j$ and $cv_k$ that are on different phases of $\phi_j \in \phi$ and $\phi_k \in \phi$ ($\phi_j \neq \phi_k$), the following constraints cover all the possible situations with just enough safety gap between the vehicles; here $\vee$ is the OR operator:

	   \begin{equation} 
	   \begin{split}
	   & t_{access,j} - t_{access,k} \geq t_{gap2} \\[-1pt]
	   & \quad \vee \\[-4pt]
	   & t_{access,k} - t_{access,j} \geq t_{gap2} \\
	   & \ \quad \quad cv_j, cv_k \in CV; \\[-1pt]
	   & \ \quad \quad \phi_j, \phi_k \in \phi, \quad \phi_j \neq \phi_k.
	   \label{eqn:constraint_tgap2_rephrased}
	   \end{split}
	   \end{equation}

\noindent where $t_{gap2}$ is the safety gap we need between access times which would be the longest it takes for a vehicle to pass the intersection area from access point to exit point. For an average acceleration of 2 m/s$^2$, $t_{gap2}$ is 7.5 sec.

	\begin{figure}
	%\centering
	\setlength\fboxsep{0pt}
	\setlength\fboxrule{0pt}
	\fbox{\includegraphics[width=\columnwidth, trim={0 0 0 0},clip]{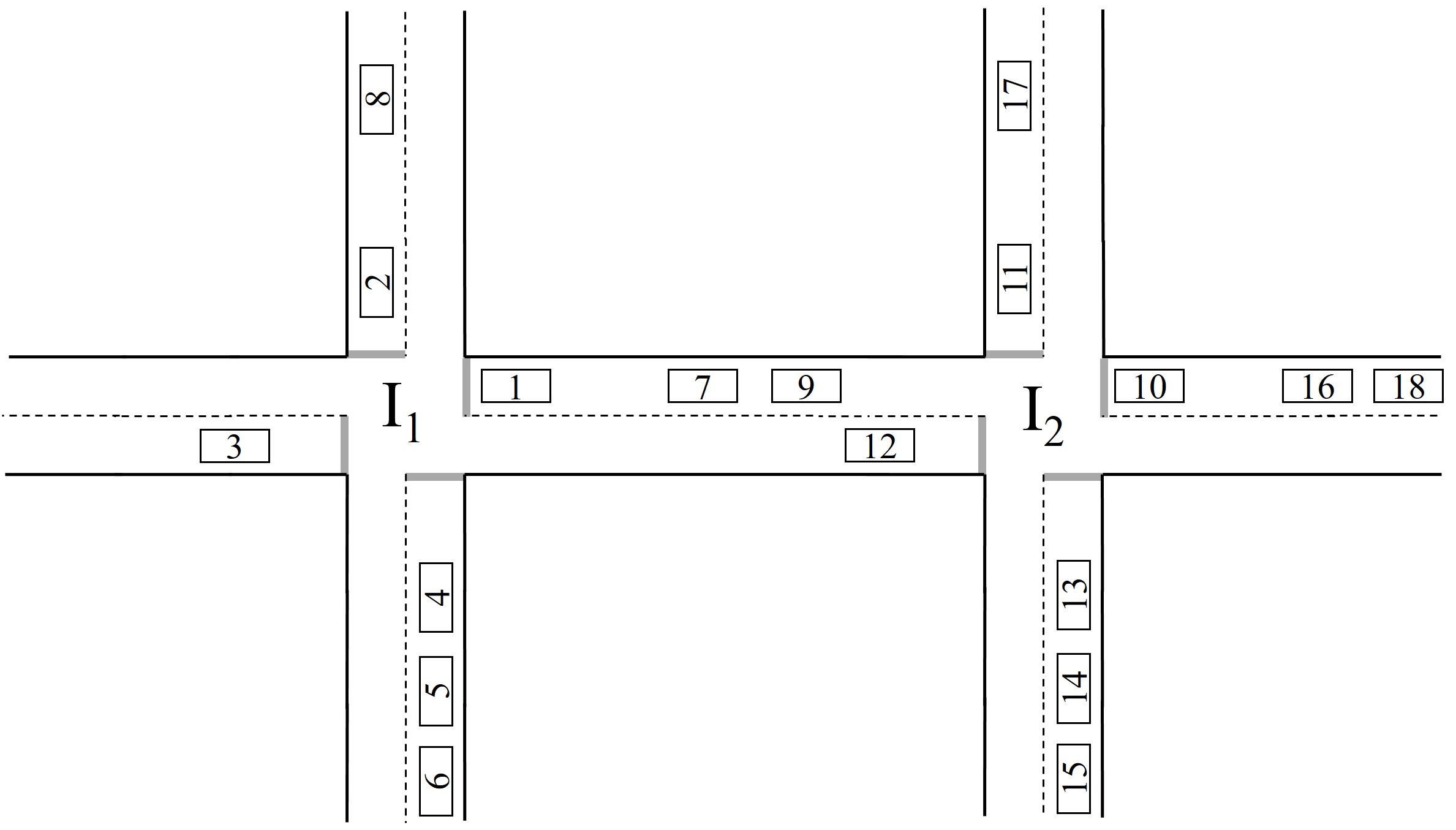}}
	\caption{\footnotesize MATLAB case study schematic. Vehicles numbered from 1 to 9 subscribe to the first intersection ($I_1$) and 10 to 18 subscribe to the second intersection ($I_2$).}
	\label{fig:MILP_schematic_matlab}
	\end{figure}
  
%  ===========================================================================
\section{Distributed Multi-Intersection Optimization}
\label{multi_int}
We apply the scheduling formulation in Section \ref{sig_control} to a grid of intersections. In our simulation, we consider a 3 by 3 grid in which there is one intersection with all four phases joined to neighboring intersections.

We evaluate two methods for implementing the MILP formulation in a grid intersection topology. In the first method, each intersection only considers vehicles subscribed to it and does not pass information about those vehicles to neighboring intersections.

In the second method, we will also add a scalable coordination scheme between multiple intersection controllers in the grid. Effective passage of a group of vehicles requires real-time coordination of neighboring intersection controllers. In the proposed distributed control approach, each intersection processes information from its subscribed vehicles, optimizes their access times, and then passes on the list of subscribed vehicles to the neighboring intersections along with their anticipated desired access times to those neighboring intersections. The desired access time of each vehicle to the second intersection will be calculated as follows: 

\begin{equation}
  t_{access,des,i}(j) = t_{access,i}(k) + t_{travel,i}(k->j)  \label{eqn:multi_des_t}
\end{equation}

\noindent where $t_{access,des,i}(j)$ is desired access time of $cv_i$ to intersection $j$, $t_{access,i}(k)$ is the vehicle's access time to the previous intersection $k$, and $t_{travel,i}(k->j)$ is the desired travel time for the vehicle between the two intersections.
%  ===========================================================================
\section{MILP Case Study} 
\label{p2_ch:signalControl_sec:milp}
In this section, in order to illustrate the effect of communication between intersections, our MILP formulation were applied to a simplified grid which consists of only two intersections. Later we show the results for a 3 by 3 grid. 

To solve our illustrative case study in this section, we use the \textit{intlinprog} function the Optimization Toolbox of Matlab (version R2016a). % The steps are stated as follows [29]: %Optimization toolbox [21]. This function solves problems in the form:
We simulate $n$=9 connected vehicles, $cv_i$ (1 $\leq$ i $\leq$ 9) at the first intersection and (10 $\leq$ i $\leq$ 18)at the second intersection. We set the speed limit to $v_{max}$=72.4 kph (45 mph), and average arterial road speed to $v_{avg}$=56.3 kph (35 mph). We assume that the current state of all vehicles is available: they are all traveling at $v_{avg}$ and their initial distances to the access area are [690 m, 750 m, 780 m, 900 m, 990 m, 1080 m, 1170 m, 1230 m, 1290 m], respectively. The distance between the two intersections is 500 m. Note that the vehicles' initial states in the first intersection are identical to those in the second intersection. 

Figure \ref{fig:MILP_schematic_matlab} demonstrates a schematic of the two-intersection grid used in our case study. For each intersection, nine vehicles initially subscribe. Vehicles number 1 to 9 subscribe to $I_1$ and 10 to 18 to $I_2$. Five of the subscribed vehicles to each intersection are traveling on phase X and four on phase O. No turn is allowed and some of the vehicles on phase O of both intersections are traveling to their neighboring intersection. In the first MILP iteration, each of the nine vehicles subscribed to their intersection will be assigned an access time. Then, in the second iteration, all vehicles on phase O will subscribe to both intersections according to Equation (\ref{eqn:multi_des_t}). As represented in Table \ref{tbl:MATLAB_results}, some of the vehicles' access times to their first intersection, are changed due to subscription of vehicles to the neighboring intersection. In microsimulation we do not need to include an iterative process. The MILP engine runs every 6 seconds so intersections start to pass on information about subscribing vehicles to their neighbors after the the first run.

 	\begin{table}       
			 \captionsetup{width= 8.5 cm} 	         
 	         \caption{\vspace{-0.5mm}\footnotesize Vehicles access times in MATLAB case study obtained by considering 500 m distance between two intersections. As it can be seen in Figure \ref{fig:MILP_schematic_matlab}, vehicles $cv_3$, $cv_{10}$, $cv_{16}$, $cv_{18}$ pass both intersections and their access times are represented in the table. }
 	         \label{tbl:MATLAB_results}
 	%\centering
 		%\renewcommand{\arraystretch}{0.6}% Tighter line spacing
 	 	 	\small %\scriptsize %\small
 	    \begin{tabular}{c c c|c c}
 	  %  \hline\hline
\toprule
      \textbf{ } & \multicolumn{2}{c}{\textbf{First Iteration}} & \multicolumn{2}{c}{\textbf{Second Iteration}}\\
\midrule
      % %
      \textbf{Vehicle ID} & \textbf{$I_1$} & \textbf{$I_2$}& \textbf{$I_1$} & \textbf{$I_2$}\\
      % %
\midrule
         1 	& 44.10 & - & 44.10 & - \\
         2 	& 51.60 & - & 51.60 & -\\
         3 	& 44.10 & - & 44.10 & 72.78\\
         4 	& 57.52 & - & 57.52 & -\\
         5 	& 60.31 & - & 63.27 & -\\
         6 	& 61.31 & - & 67.28 & -\\
         7 	& 68.81 & - & 74.78 & -\\
         8 	& 61.31 & - & 84.27 & -\\
         9 	& 69.81 & - & 75.78 & -\\
         10 & - & 44.10 & 76.78 & 44.10\\
         11 & - & 51.60 & - & 51.60\\
         12 & - & 44.10 & - & 44.10\\
         13 & - & 57.52 & - & 57.52\\
         14 & - & 60.31 & - & 63.27\\
         15 & - & 61.31 & - & 64.27\\
         16 & - & 68.81 & 96.80 & 71.77\\
         17 & - & 61.31 & - & 64.27\\
         18 & - & 68.81 & 97.80 & 72.78\\	   
\bottomrule    
    \end{tabular}
\end{table}

\section{Benchmarking}
\label{bench}
%The key trait of all the traffic control projects is to be more, in general word, better than the current or other proposed traffic management techniques. Therefore, there should be other benchmark testbeds for this analogy.
We compare our proposed distributed MILP intersection control testbed to two fixed benchmark testbeds and also to a MILP-based intersection control without coordination. The first testbed (Testbed A) has all fixed-time intersections. The second testbed (Testbed B) is also fixed-time; with the difference that each vehicle adjusts its trajectory individually for arrival at green based on prior knowledge of traffic signal state (see \cite{12AsadiB.2011, fayazi2018IEEE}). The fixed signal timing of each intersection was obtained from Trafficware's Synchro Studio software which calculates optimal fixed timings for each intersection in the grid based on the flow of vehicles. Testbed C is MILP-based but lacks coordination between intersections. Vehicles are injected to the grid in a same pattern for all testbeds but the injection rate varies at each boundary node of the  grid. 
% % % % % % % % % % % % % % % % % % % % % % % % % % % % % % % %
% % % % % % % % % % % % % % % % % % % % % % % % % % % % % % % %
% % % % % % % % % % % % % % % % % % % % % % % % % % % % % % % %
\subsection{Testbed A: Pre-timed Signalized Intersection Grid}
\label{p2_ch:testResults_sec:benchmarks_subsec:fixedSemiInfo}

In this testbed, optimized fixed signal timing has been used for each intersection in the grid. It is assumed that autonomous vehicles cameras can detect signal status in a 300 m range. Vehicles plan their velocity and acceleration by ``watching'' the signal ahead of them, and given constraints imposed by surrounding traffic. More detailed explanation of this testbed can be found in \cite{fayazi2018IEEE}. %An additional constraint on the intersections common phases is added to the simulation to make sure the phases are not overfull of vehicles. For instance, in \ref{fig:MILP_schematic_matlab} the vehicle on phase $O''$ of the first intersection will go to phase $O''$ of the second intersection after it passes the first intersection. However, if the second intersection phase $O''$ is full, then the vehicle should stay behind the first intersection stop bar.

% % % % % % % % % % % % % % % % % % % % % % % % % % % % % % % %
% % % % % % % % % % % % % % % % % % % % % % % % % % % % % % % %
% % % % % % % % % % % % % % % % % % % % % % % % % % % % % % % %
\subsection{Testbed B: Communication-based Trajectory-Planning at Pre-timed Signalized Intersection Grid}
\label{p2_ch:testResults_sec:benchmarks_subsec:fixedFullInfo}

In this testbed, vehicles will be provided with deterministic future state of traffic signals via unidirectional wireless communication when they are within 400 m of the intersection. The autonomous vehicles plan their velocity and acceleration trajectory based on the speed advisory algorithm proposed by our group in \cite{12AsadiB.2011}. Besides access to signal timing data, vehicles receive queue size information of the intersection they are traveling to, when they are 300 m behind the intersection stop bar.

% % % % % % % % % % % % % % % % % % % % % % % % % % % % % % % %
% % % % % % % % % % % % % % % % % % % % % % % % % % % % % % % %
% % % % % % % % % % % % % % % % % % % % % % % % % % % % % % % %
\subsection{Testbed C: MILP-based Trajectory-Planning without intersection coordination}
\label{p2_ch:testResults_sec:benchmarks_subsec:milpnocoInfo}

In this testbed, each intersection is equipped with a MILP controller and traffic signals are removed. Also, vehicles are assumed to have an automatic velocity control. Each intersection keeps information about the subscribed vehicles to itself and does not share it with neighboring intersections. Therefore, each intersection solves its own optimization problem regardless of what is happening in other intersections.

%	\begin{figure}[h]
%	\centering
%	\includegraphics[width=\columnwidth, trim={1.3cm 4cm 2cm 4cm},clip]{VAS.pdf}
%	\caption{\footnotesize The graphics shows how our modified speed adisory algorithm finds the feasible velocity intervals in order to avvoid stopping at red, if possible.}
%	\label{fig:VAS}
%	\end{figure}

 	\begin{table*}        
 	         \caption{\vspace{-0.5mm}\footnotesize SIL simulation results, and the overall performance improvements achieved by MILP-based multi-intersection controller (Testbed Coordinated MILP)}
 	         \label{tbl:SILs_results}
 	\centering
 		\renewcommand{\arraystretch}{0.6}% Tighter line spacing
 	 	 	\small %\scriptsize %\small
 	    \begin{tabular}{c|c|c|c|c}
      \textbf{MOE} & \textbf{Testbed} & \textbf{Testbed}& \textbf{Testbed} & \textbf{Testbed} \\
      % %
      \textbf{(for all simulated vehicles)} & \textbf{A} & \textbf{B}& \textbf{C} & \textbf{Coordinated MILP}\\
      % %
\midrule
         Intersection traversals 	& 5802 & 5804 & 5803 & 5801\\
\midrule
          NO. vehicles made their destination	& 5526 & 5518 & 5574 & 5566\\
\midrule
		 Total number of stops & 6886 & 5687 & 2176 & 2140\\   
\midrule
	  	Average stop time per each vehicle & 23sec & 17sec & 7sec & 8sec \\   
\midrule
	   Average travel time per vehicle & 2min 40sec & 2min 45sec & 2min 2sec  & 1min 54sec \\\midrule
	   Average miles per gallon per vehicle & 26.3 mpg & 26.7 mpg  & 28.6 mpg & 30.7 mpg \\  
\midrule   	   
    \end{tabular}
    %\vspace{-1mm}
\end{table*}

%  ===========================================================================
\section{Microsimulation}
\label{p2_ch:signalControl_sec:SIL}
\begin{figure}
	\centering
	\setlength\fboxsep{0pt}
	\setlength\fboxrule{0pt}
	\fbox{\includegraphics[width=6.2cm, trim={0 0 0 0},clip]{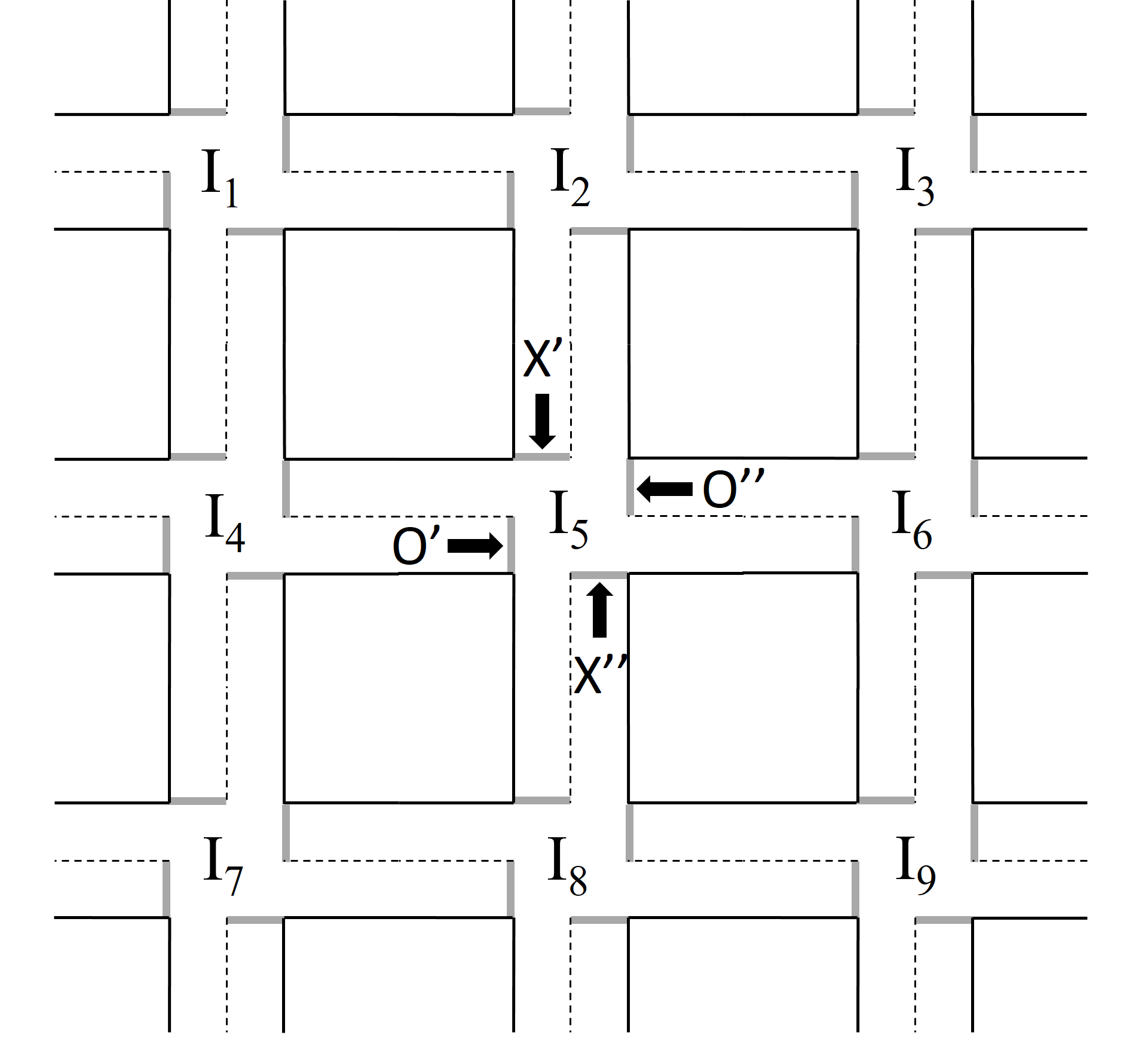}}
	\caption{\footnotesize Schematic of a 3 by 3 intersection grid. Intersections are labeled as $I_i$ where $i$ is the intersection number. All phases of the middle intersection $I_5$ are labeled according to the notation presented in Figure \ref{fig:intersection_system}.}
	\label{fig:Grid_schematic}
	\end{figure}

We implemented a software-in-the-loop (SIL) microsimulation using the Java programming language. The 3 by 3 intersection grid architecture is presented in Figure \ref{fig:Grid_schematic}. 

Vehicles are injected to the grid from 12 links on the perimeter of the grid at [550, 300, 600, 950, 550, 200, 750, 900, 400, 450, 750] vehicles per hour, which is randomly generated. The injection rate is held constant for all testbeds. The length of each intersection phase in the grid is 400 m. Therefore, each vehicle will subscribe to its destination intersection when it reaches the 400 m range. Speed limit is set to $v_{max}$=72.4 kph (45 mph) and average velocity is set to $v_{avg}$=56.3 kph (35 mph). IBM's CPLEX optimization package is used to solve the grid's MILP formulation and the simulation was run for 1 hour in all testbeds. The controller solves the formulated optimization problem every 6 seconds to provide the vehicles with a real-time feedback. From our MATLAB case study presented in Section \ref{p2_ch:signalControl_sec:milp} two iterations were needed for each intersection to receive full list of subscribing vehicles. However, it might not be computationally cost effective to solve the MILP problem two times per step per intersection in this larger scale microsimulation. Therefore, we run the MILP engine once per step per intersection but reduce the step time to reduce the error. Note that the information from the neighboring intersections will be one step old in this scenario. 

A fuel economy analysis has been done using the data and model presented in \cite{Austin2017ITSC}. Stored velocity and acceleration data of each vehicle from our microsimulation were used for this analysis. The model in \cite{Austin2017ITSC} includes drag, gravitational and friction forces in order to calculate the total traction force in the wheels. Optimal gear selection contours at vehicle operation point space is derived as a function of vehicle acceleration and velocity by minimizing the fuel mass flow rate at every operating point. Therefore, we can find the optimal gear selection at every point in time having the velocity and acceleration of the vehicle. Then, engine torque and rpm are determined from which the fuel consumption rate can be calculated.
 
Grid level measures of effectiveness are presented in Table \ref{tbl:SILs_results}. It can be seen that the MILP approach (Testbed C and Testbed Coordinated MILP) improves travel time, stop time, and number of stops significantly compared to fixed time testbeds. An interesting observation is the small difference between testbed A (conventional fixed time) and testbed B (fixed time with speed advisory). Previously in \cite{fayazi2017ACC}, the difference between the two testbeds' number of stops and idling delay were significant enough to conclude that speed advisory is very efficient. That is because in \cite{fayazi2017ACC} vehicles received speed advisory 2 km prior to intersection and could plan their velocity trajectory well in advance. In this paper, the length of all links is 400 m so signal detection range of 400 m in testbed B is not significantly higher than the assumed 300 m for testbed A. This weakens the performance of speed advisory.

Fuel economy analysis shows the two MILP based testbeds are significantly increasing the average miles per gallon (MPG) per vehicle. Considering average MPG per vehicle in testbed A as a reference for comparison, an improvement of 1.5\%, 8.7\%, and 16.7\% can be seen in Testbed B, Testbed C, and Testbed Coordinated MILP, respectively. The considerable higher average MPG in Testbed Coordinated MILP compared to Testbed C is because of the lower acceleration and deceleration of the vehicles in coordinated scenario. This can be the most effective role of coordination between intersections. When each vehicle is informed of its access time to two consecutive intersections, it would plan its velocity well in advance to avoid high acceleration and deceleration.
\begin{figure}
	\centering
	\setlength\fboxsep{0pt}
	\setlength\fboxrule{0pt}
	\fbox{\includegraphics[width=8.1cm, trim={0 0 0 0},clip]{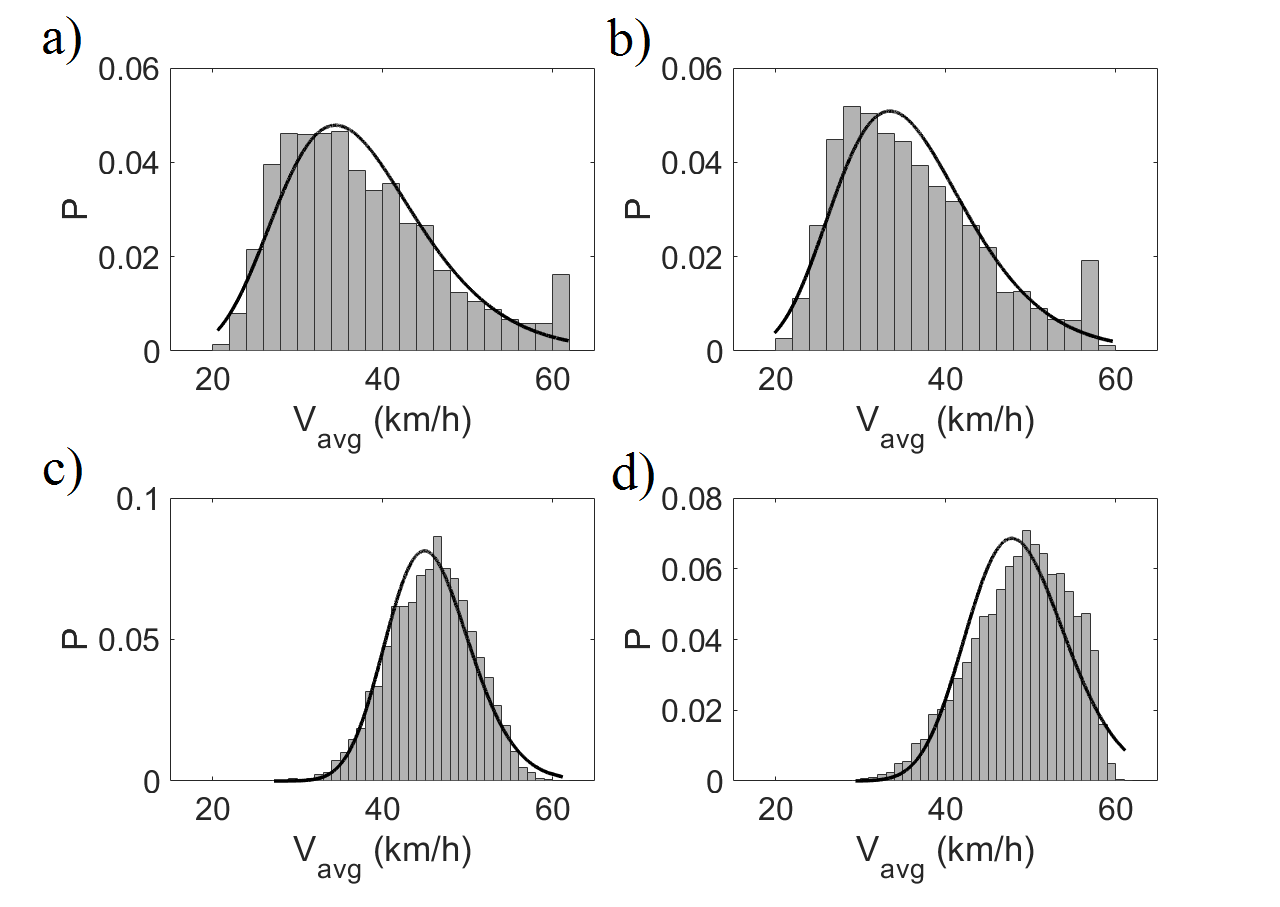}}
	\caption{\footnotesize Histograms and lognormal fits of average velocity of vehicles in all four testbeds a) Testbed A, b) Testbed B, c) Testbed C, d) Coordinated MILP.}
	\label{fig:Velocity_distribution}
	\end{figure}

A statistical analysis on average velocity of vehicles also shows the effectiveness of MILP intersection control. In Figure \ref{fig:Velocity_distribution} the mean velocity in coordinated MILP is higher than other testbeds. Moreover, both coordinated and uncoordinated MILP velocity lognormal distribution are slimmer than others which means there are less fluctuations in vehicles velocity in these two testbeds. One can infer from this fact that vehicles do not accelerate and decelerate as much as in other cases, which improves the fuel efficiency.

\begin{figure}
	\centering
	\setlength\fboxsep{0pt}
	\setlength\fboxrule{0pt}
	\fbox{\includegraphics[width=8.5cm, trim={0 0 0 0},clip]{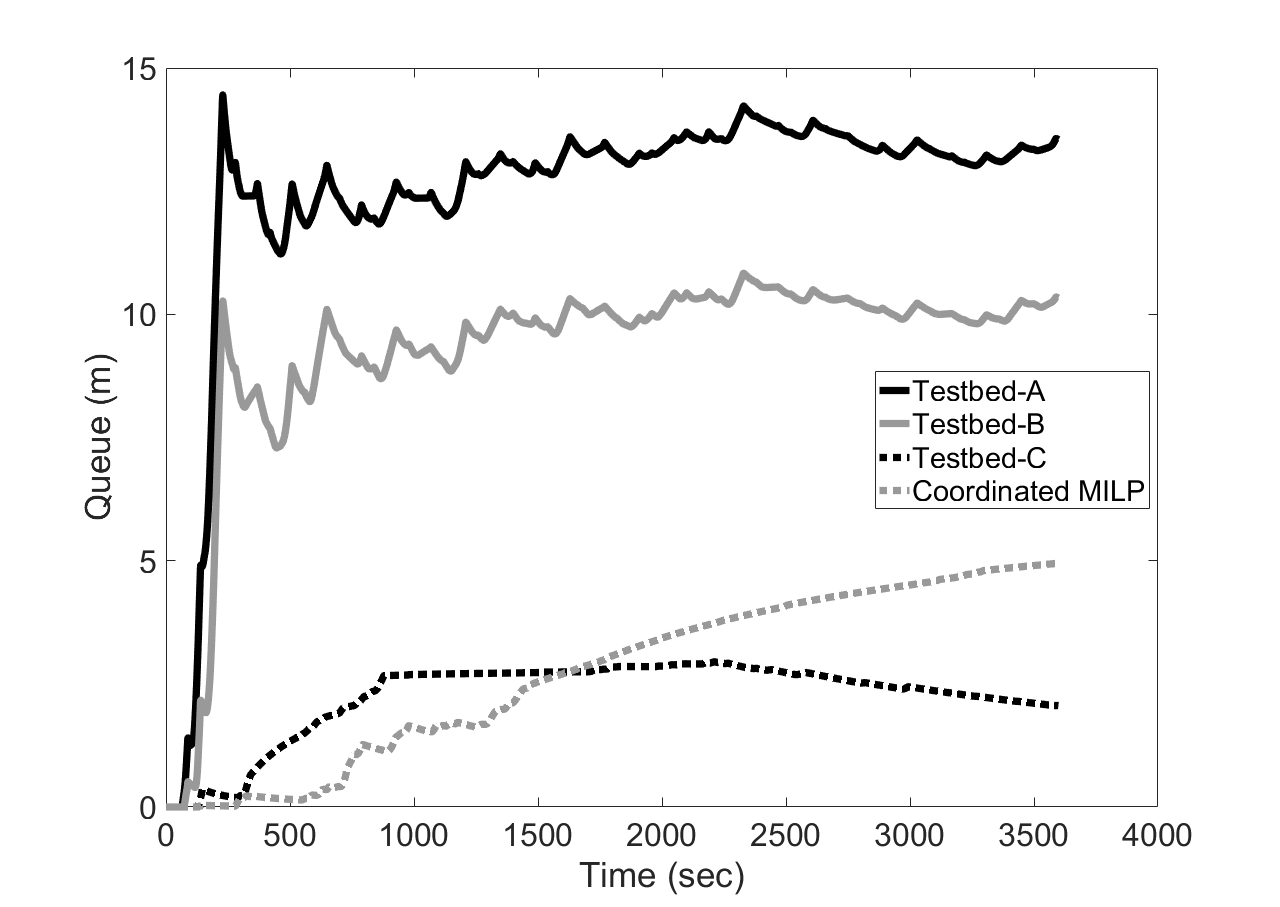}}
	\caption{\footnotesize Averaged queue of all phase of $I_5$ from all four testbeds. It can be seen that average queue in MILP testbeds are significantly lower than testbeds A and B.}
	\label{fig:Average_queue}
	\end{figure}

We are specifically interested in the middle intersection denoted by $I_5$ in Figure \ref{fig:Grid_schematic} since it communicates with four neighboring intersections. The plot of averaged queue size of 4 phases of this intersection can be seen in Figure \ref{fig:Average_queue}. The average queue size is significantly lower in MILP based control methods (Testbed C and Testbed Coordinated MILP) compared to testbeds A and B.

%
%
%\vspace{1mm}
\section{Conclusions}
In this paper, the mixed integer linear programming intersection control method previously published in \cite{fayazi2017ACC, fayazi2017CCTA, fayazi2018IEEE} was applied to a grid of 9 intersections. A simple case study involving only two intersections was studied first to examine the intersections' communication. Then, a microsimulation was performed on a 3 by 3 intersection grid. We had two multi-intersection testbeds . One with coordination between intersection controllers and the other one without such coordination. Our results show that the distributed MILP control methods significantly reduce stop time, travel time, fuel consumption rate, and number of stops compared to other testbeds with fixed-time traffic signals. In coordinated testbed, vehicles travel with higher average velocities compared to that of uncoordinated testbed (Testbed C) which result in a lower average travel time at the expense of higher average stop delay per each stop. The most significant contribution of the coordination scheme is to help vehicles avoid high acceleration/deceleration. This will result in a substantial improvement in average miles per gallon (MPG) per vehicle in coordinated testbed (Testbed Coordinated MILP) compared to uncoordinated testbed (Testbed C). The better fuel economy is acheived without incorporating it into our optimal control formulation. In contrast to our previous results, the speed advisory method that is based on prior knowledge of traffic signal state is not significantly better than fixed time signal traffic control. This can be improved by providing future state of consecutive traffic signals to vehicles. Potential future works include considering turning of vehicles, multiple lane roads, using geographical data to simulate a real intersection grid in a city, and incorporating fuel economy analysis. Dedicated Short Range Communication (DSRC) offers wireless connectivity for traffic elements within the required distances for our intersection architecture \cite{kenney2011dedicated}. Other future work could apply it as part of an experimental vehicle-in-the-loop implementation.

%\section*{APPENDIX}

\section*{ACKNOWLEDGMENT}
Authors would like to thank R. Austin Dollar for his contribution in fuel economy analysis.

%%%%%%%%%%%%%%%%%%%%%%%%%%%%%%%%%%%%%%%%%%%%%%%%%%%%%%%%%%%%%%%%%%%%%%%%%%%%%%%%
%
\bibliographystyle{ieeetr}
\bibliography{root}
\end{document}